\documentclass[runningheads]{llncs}
\usepackage[T1]{fontenc}
\usepackage{graphicx}
\usepackage{bm}
\usepackage{multirow}
\usepackage{hyperref}
\usepackage{amsfonts,amssymb,amsmath}
\usepackage{pifont}
\usepackage{epstopdf}
\usepackage{color}
\usepackage{cite}
\newcommand{\cmark}{\ding{51}}%
\newcommand{\xmark}{\ding{55}}%
\usepackage[misc]{ifsym}
\usepackage{subfigure}
\usepackage{amsmath,amssymb}
\usepackage{makecell}
\usepackage{subfigure}
\usepackage[misc]{ifsym}
\usepackage{xcolor}
\usepackage{multirow}
\usepackage{algorithm}
\usepackage{algorithmic}

\begin{document}
\title{Federated Uncertainty-Aware Aggregation\\ for Fundus Diabetic Retinopathy Staging}
\author{Meng Wang\inst{1\#}\and
Lianyu Wang\inst{2\#} \and
Xinxing Xu\inst{1}\and
Ke Zou\inst{3}\and
Yiming Qian\inst{1}\and \\
Rick Siow Mong Goh\inst{1}\and
Yong Liu\inst{1}\and
Huazhu Fu\inst{1 (\textrm{\Letter})}
}
\titlerunning{FedUAA}

\authorrunning{M. Wang et al.}
%
\institute{Institute of High Performance Computing (IHPC), Agency for Science, Technology and Research (A*STAR), 1 Fusionopolis Way, \#16-16 Connexis, Singapore 138632, Republic of Singapore \and
College of Computer Science and Technology, Nanjing University of Aeronautics and Astronautics, Jiangsu 211100, China.
\and
National Key Laboratory of Fundamental Science on Synthetic Vision and the College of Computer Science, Sichuan University, Sichuan 610065, China.\\
\# Meng Wang and Lianyu Wang contributed equally.\\
\textrm{\Letter} Corresponding author: Huazhu Fu (\email{hzfu@ieee.org})
}
\maketitle              
\begin{abstract}
Deep learning models have shown promising performance in the field of diabetic retinopathy (DR) staging. However, collaboratively training a DR staging model across multiple institutions remains a challenge due to non-iid data, client reliability, and confidence evaluation of the prediction. To address these issues, we propose a novel federated uncertainty-aware aggregation paradigm (FedUAA), which considers the reliability of each client and produces a confidence estimation for the DR staging.
In our FedUAA, an aggregated encoder is shared by all clients for learning a global representation of fundus images, while a novel temperature-warmed uncertainty head (TWEU) is utilized for each client for local personalized staging criteria. Our TWEU employs an evidential deep layer to produce the uncertainty score with the DR staging results for client reliability evaluation.
Furthermore, we developed a novel uncertainty-aware weighting module (UAW) to dynamically adjust the weights of model aggregation based on the uncertainty score distribution of each client.
In our experiments, we collect five publicly available datasets from different institutions to conduct a dataset for federated DR staging to satisfy the real non-iid condition. The experimental results demonstrate that our FedUAA achieves better DR staging performance with higher reliability compared to other federated learning methods.
Our proposed FedUAA paradigm effectively addresses the challenges of collaboratively training DR staging models across multiple institutions, and provides a robust and reliable solution for the deployment of DR diagnosis models in real-world clinical scenarios.

\end{abstract}
\section{Introduction}
In the past decade, numerous deep learning-based methods for DR staging have been explored and achieved promising results~\cite{Gulshan2016,Ting2017,gunasekeran2020artificial,fundus_survey}.
However, most current studies focus on centralized learning, which necessitates data collection from multiple institutions to a central server for model training.
This approach poses significant data privacy security risks.
Additionally, in clinical practice, different institutions may have their own DR staging criteria~\cite{asiri2019deep}. 
Consequently, it is difficult for the previous centralized DR staging method to utilize data of varying DR staging criteria to train a unified model.
\vspace{-0.9pt}

Federated learning (FL) is a collaborative learning framework that enables training a model without sharing data between institutions, thereby ensuring data privacy~\cite{Li2020_FL,Kairouz2021}. In the FL paradigm, FedAvg~\cite{FedAvg} and its variants~\cite{FedRep,Fed_SCAFFOLD,FedDyn,FedMoon,FedProx,FedDC,FedBN} are widely used and have achieved excellent performance in various medical tasks. However, these FL methods assign each client a static weight for model aggregation, which may lead to the global model not learning sufficient knowledge from clients with large heterogeneous features and ignoring the reliability of each client.
In clinical practice, the data distributions of DR datasets between institutions often vary significantly due to medical resource constraints, population distributions, collection devices, and morbidity~\cite{diagnostics12112835,10.1007/978-3-031-16449-1_65}. This variation poses great challenges for the exploration of federated DR staging methods. Moreover, most existing DR staging methods and FL paradigms mainly focus on performance improvement and ignore the exploration of the confidence of the prediction. Therefore, it is essential to develop a new FL paradigm that can provide reliable DR staging results while maintaining higher performance. Such a paradigm would reduce data privacy risks and increase user confidence in AI-based DR staging systems deployed in real-world clinical settings.

To address the issues, we propose \textbf{a novel FL paradigm, named FedUAA}, that employs a personalized structure to handle collaborative DR staging among multiple institutions with varying DR staging criteria. We utilize uncertainty to evaluate the reliability of each client's contribution. While uncertainty is a proposed measure to evaluate the reliability of model predictions~\cite{TMC,huang2022evidence,zou2022tbrats,yu2022robust}, it remains an open topic in FL research. In our work, we introduce \textbf{a temperature-warmed evidential uncertainty (TWEU)} head to enable the model to generate a final result with uncertainty evaluation without sacrificing performance. Additionally, based on client uncertainty, we developed \textbf{an uncertainty-aware weighting module (UAW)} to dynamically aggregate models according to each client's uncertainty score distribution. This can improve collaborative DR staging across multiple institutions, particularly for clients with large data heterogeneity. 
Finally, we construct a \textbf{dataset for federated DR staging} based on five public datasets with different staging criteria from various institutions to satisfy the real non-iid condition.
The comprehensive experiments demonstrate that FedUAA provides outstanding DR staging performance with a high degree of reliability, outperforming other state-of-the-art FL approaches. 
\section{Methodology}
\begin{figure}[!t]
\centering
\includegraphics[width=0.9\textwidth]{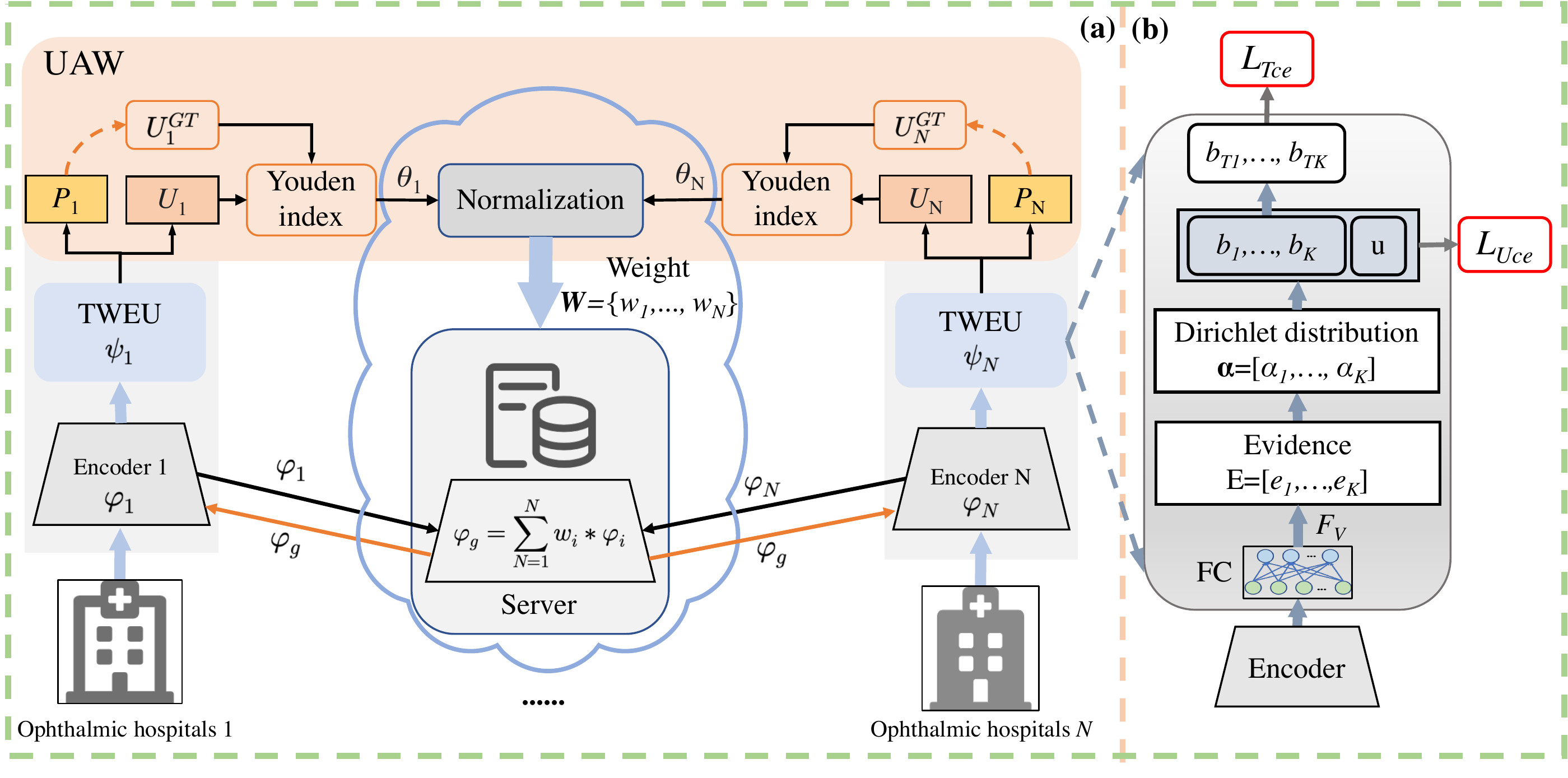}
\vskip -5pt
\caption{The overview of FedUAA (a) with  TWEU module (b). An aggregated encoder is shared by all clients for learning a global representation of fundus images, while a novel TWEU head is kept on the local client for local personalized staging criteria. Furthermore, a novel UAW module is developed to dynamically adjust the weights for model aggregation based on the reliability of each client. 
}
\label{Framework}
\end{figure}

Fig.~\ref{Framework}~(a) shows the overview of our proposed FedUAA. During training, local clients share the encoder~($\varphi$) to the cloud server for model aggregation, while the TWEU~($\psi$) head is retained locally to generate DR staging results with uncertainty evaluation based on features from the encoder to satisfy local-specific DR staging criteria. The algorithm of our proposed FedUAA is detailed in \textbf{Supplementary A}.
Therefore, the target of our FedUAA is:
\begin{equation}
\min_{\varphi \in \Phi ,\psi \in \Psi } \sum^{N}_{i=1} \pounds \left( f_{i}\left( \varphi_{i} ,\psi_{i} |X_{i}\right),Y_{i}\right),  
\label{eq:1}
\end{equation}
where $\pounds$ is the total loss for optimizing the model, $f_{i}$ is the model of \textit{i}-th client, while $X_{i}$ and $Y_{i}$ are the input and label of \textit{i}-th client.
Different from previous personalized FL paradigms~\cite{arivazhagan2019federated,FedRep}, our FedUAA dynamically adjusts the weights for model aggregation according to the reliability of each client,
i.e., the client with larger distributional heterogeneity tends to have larger uncertainty distribution and should be assigned a larger weight for model aggregation to strengthen attention on the client with data heterogeneity.
Besides, by introducing TWEU, our FedUAA can generate a reliable prediction with an estimated uncertainty, which makes the model more reliable without losing DR staging performance.

\subsection{Temperature-warmed evidential uncertainty head}
\label{TWEU}
To make the model more reliable without sacrificing DR staging performance, we propose a novel temperature-warmed evidence uncertainty head~(TWEU), which can directly generate DR staging results with uncertainty score based on the features from the encoder.
The framework of TWEU is illustrated in Fig.~\ref{Framework}~(b).
Specifically, we take one of the client models as an example and we assume that the staging criteria of this client is \textit{K} categories.
Correspondingly, given a color fundus image input, we can obtain its \textit{K}+1 non-negative mass values, whose sum is 1. This can be defined as $\sum^{K}_{i=1} b_{i}+u=1$,
where $b_{i}\geq0$ is the probability of \textit{i}-th category, while $u$ represent the overall uncertainty score. 
Specifically, as shown in Fig.~\ref{Framework}~(b), a local fully connected layer~(FC) is used to learn the local DR category-related features $F_{V}$, and the \textit{Softplus} activation function is adopted to obtain the evidence $E = [e_{1},...,e_{K}]$ of \textit{K} staging categories based on $F_{V}$, so as to ensure that its feature value is greater than 0.
Then, $E$ is re-parameterized by Dirichlet concentration~\cite{connor1969concepts}, as:
$\bm{\alpha} =E+1,\ i.e,\ \alpha_{k} =e_{k}+1$
where $\alpha_{k}$ and $e_{k}$ are the \textit{k}-th category Dirichlet distribution parameters and evidence, respectively.
Further calculating the belief masses~($\bm{b}$) and corresponding uncertainty score~($u$) by $b_{k}=\frac{e_{k}}{S} =\frac{\alpha_{k}-1}{S}, \;  u=\frac{K}{S}$,
where $S=\sum\nolimits^{K}_{k=1} \alpha^{k}_{i,j}$ is the Dirichlet intensities.
Therefore, the probability assigned to category \textit{k} is proportional to the observed evidence for category \textit{k}. Conversely, if less total evidence is obtained, the greater the uncertainty score will be.
As shown in Fig.~\ref{Framework}~(b), $L_{Uce}$ is used to guide the model optimization based on the belief masses~($\bm{b}$) and their corresponding uncertainty score~($u$).
Finally, temperature coefficients~$\tau$ is introduced to further enhance the classifier's confidence in belief masses, i.e, $b_{Ti}=\frac{e^{\left( b_{i}/\tau \right)  }}{\sum\nolimits^{K}_{i=1} e^{\left( b_{i}/\tau \right)}}$,
where $\bm{b_{T}}=\left[ b_{T1},...,b_{Tk}\right]$ is the belief masses that were temperature-warmed. As shown in Fig.~\ref{Framework}~(b), $L_{Tce}$ is adopted to guide the model optimization based on the temperature-warmed belief features of $\bm{b_{T}}$.
\subsection{Uncertainty-aware weighting module}
Most existing FL paradigms aggregate model parameters by assigning a fixed weight to each client, resulting in limited performance on those clients with large heterogeneity in their data distributions.
To address this issue, as shown in Fig.~\ref{Framework}~(a), we propose a novel uncertainty-aware weighting~(UAW)~module that can dynamically adjust the weights for model aggregation based on the reliability of each client, which enables the model to better leverage the knowledge from different clients and further improve the DR staging performance.
Specifically, at the end of a training epoch, each client-side model produces an uncertainty value distribution~($U$), and the ground truth for incorrect prediction of $U^{GT}$ also can be calculated based on the final prediction $P$ by,
\begin{equation} 
u^{GT}_{i}=1-\bm{1}\left\{ P_{i},Y_{i}\right\} , 
\; \text{where} \; 
\bm{1} \left\{ P_{i},Y_{i}\right\}  =\begin{cases}1& \text{if}\  P_{i}=Y_{i}\\ 0& \text{otherwise}\end{cases}  ,
\label{eq:8}
\end{equation}
where $P_{i}$ and $Y_{i}$ are the final prediction result and ground truth of \textit{i}-th sample in local dataset.
Based on $U$ and $U^{GT}$, we can find the optimal uncertainty score $\theta$, which can well reflect the reliability of the local client.
To this end, we calculate the ROC curve between $U$ and $U^{GT}$, and obtain all possible sensitivity~($Sens$) and specificity~($Spes$) values corresponding to each uncertainty score~($u$) used as a threshold.
Then, Youden index~($J$)~\cite{fluss2005estimation} is adopted to obtain the optimal uncertainty score $\theta$ by:
\begin{equation} 
 \theta =\arg \max_{u} J\left( u\right)  ,\; \text{with} \; \; J\left( u\right)  =Sens\left( u\right)  +Spes\left( u\right)-1.  
\label{eq:9}
\end{equation}
More details about Youden index are given in \textbf{Supplementary B}. 
Finally, the optimal uncertainty scores $\Theta =\left[ \theta_{1} ,...,\theta_{N} \right] $ of all clients are sent to the server, and a Softmax function is introduced to normalize $\Theta$ to obtain the weights for model aggregation as $w_{i}={e^{\theta_{i} }}  /{\sum\nolimits^{N}_{i=1} e^{\theta_{i} }}$.
Therefore, the weights for model aggregation are proportional to the optimal threshold of the client.
Generally, local dataset with larger uncertainty distributions will have a higher optimal uncertainty score $\theta$, indicating that it is necessary to improve the feature learning capacity of the client model to further enhance its confidence in the feature representation, and thus higher weights should be assigned during model aggregation.

\section{Loss function}
\label{Loss}
As shown in Fig.~\ref{Framework}~(b), the loss function of client model is:
\begin{equation} 
L=L_{Uce}+L_{Tce},  
\label{eq:12}
\end{equation}
where $L_{Uce}$ is adopted to guide the model optimization based on the features~($\bm{b}$ and $u$) which were parameterized by Dirichlet concentration. Given the evidence of $E=\left[ e_{1},...,e_{k}\right]$, we can obtain Dirichlet distribution parameter $\bm{\alpha}=E+1$, category related belief mass $\bm{b}=\left[ b_{1},...,b_{k}\right]$ and uncertainty score of $u$. Therefore, the original cross-entropy loss is improved as,
\begin{equation} 
L_{Ice}\! =\! \int \left[ \sum^{K}_{k=1} -y_{k}\log \left( b_{k}\right)  \right]  \frac{1}{\beta \left( \alpha \right)  } \prod^{K}_{k=1} b^{\alpha_{k} -1}_{k}db=\sum^{K}_{k=1} y_{k}\left( \Phi \left( S\right)  -\Phi \left( \alpha_{k} \right)  \right), 
\label{eq:13}
\end{equation}
where $\Phi (\cdot)$ is the digamma function, while $\beta \left( \alpha \right)$ is the multinomial beta function for the Dirichlet concentration parameter $\alpha$. Meanwhile, the \textit{KL} divergence function is introduced to ensure that incorrect predictions will yield less evidence: 
\begin{equation} 
L_{KL} = \log \left( \dfrac{\Gamma \left( \sum^{K}_{k=1} \left( \tilde{\alpha }_{k} \right) \right)  }{\Gamma \left( K\right)  \sum^{K}_{k=1} \Gamma \left( \tilde{\alpha }_{i} \right)} \right) +\sum_{k=1}^{K} \left( \tilde{\alpha }_{k} -1\right)  \left[ \Phi \left( \tilde{\alpha }_{k} \right)  -\Phi \left( \sum_{i=1}^{K} \tilde{\alpha }_{k} \right)  \right],  
\label{eq:14}
\end{equation} 
where $\Gamma (\cdot)$ is the gamma function, while $\tilde{\alpha } =y+\left( 1-y\right) \odot \alpha$ represents the adjusted parameters of the Dirichlet distribution which aims to avoid penalizing the evidence of the ground-truth class to 0.
In summary, the loss function $L_{Uce}$ for the model optimization based on the features that were parameterized by Dirichlet concentration is as follows:
\begin{equation}
L_{Uce}=L_{Ice}+\lambda \ast L_{KL}, 
\label{eq:15}
\end{equation}
where $\lambda$ is the balance factor for $L_{KL}$. To prevent the model from focusing too much on KL divergence in the initial stage of training, causing a lack of exploration for the parameter space, 
we initialize $\lambda$ as 0 and increase it gradually to 1 with the number of training iterations. And, seen from Sec.~\ref{TWEU}, Dirichlet concentration alters the original feature distribution of $F_{v}$, which may reduce the model's confidence in the category-related evidence features, thus potentially leading to a decrease in performance.
Aiming at this problem, as shown in Fig.~\ref{Framework}~(b), we introduce temperature coefficients to enhance confidence in the belief masses,
and the loss function $L_{Tce}$ to guide the model optimization based on the temperature-warmed belief features $\bm{b_{T}}$ is formalized as: 
\begin{equation} 
L_{Tce}=-\sum^{K}_{i=1} y_{i}log\left( b_{Ti}\right).
\end{equation}

\section{Experimental results}
\label{Results}

\noindent\textbf{Dataset and Implementation:} 
We construct a database for federated DR staging based on 5 public datasets, including APTOS~(3,662 samples)~\footnote{\url{https://www.kaggle.com/datasets/mariaherrerot/aptos2019}}, Messidor~(1,200 samples)~\cite{decenciere2014feedback}, DDR~(13,673 samples)~\cite{LI2019}, KaggleDR~(35,126 samples)~(DRR)~\footnote{\url{https://www.kaggle.com/competitions/diabetic-retinopathy-detection}}, and IDRiD~(516 samples)~\cite{IDRiD}, where each dataset is regarded as a client, 
More details of datasets are given in \textbf{Supplementary C}. 

We conduct experiments on the Pytorch with 3090 GPU.
The SGD  with a learning rate of 0.01 is utilized. 
The batch size is set to 32, the number of epochs is 100, and the temperature coefficient $\tau$ is empirically set to 0.05.
To facilitate training, the images are resized to 256×256 before feeding to the model. 

\begin{table}[!t]
\centering
\caption{AUC results for different FL methods applied to DR staging.}\label{tab:ComparisonAUCs}
\begin{tabular}{l||c|c|c|c|c||c}
\hline
Methods & APTOS & DDR & DRR & Messidor & IDRiD & Average \\ \hline
SingleSet & 0.9059 & 0.8776 & 0.8072 & 0.7242 & 0.7168 & 0.8063 \\ \hline
FedRep~\cite{FedRep} & 0.9372 & 0.8964 & 0.8095 & 0.7843 & 0.8047 & 0.8464 \\ \hline
FedBN~\cite{FedBN} & 0.9335 & 0.9003 & 0.8274 & 0.7792 & 0.8193 & 0.8519 \\ \hline
FedProx~\cite{FedProx} & 0.9418 & 0.8950 & 0.8127 & 0.7877 & 0.8049 & 0.8484 \\ \hline
FedDyn~\cite{FedDyn} & 0.9352 & 0.8778 & 0.8022 & 0.7264 & 0.5996 & 0.7882 \\ \hline
SCAFFOLD~\cite{Fed_SCAFFOLD} & 0.9326 & 0.8590 & 0.7251 & 0.7288 & 0.6619 & 0.7815 \\ \hline
FedDC~\cite{FedDC} & 0.9358 & 0.8858 & 0.7969 & 0.7390 & 0.7581 & 0.8236 \\ \hline
Moon~\cite{FedMoon} & 0.9436 & 0.8995 & 0.8117 & 0.7907 & 0.8115 & 0.8514 \\ \hline
MDT~\cite{yu2022robust} & 0.9326 & 0.8908 & 0.7987 & 0.7919 & 0.7965 & 0.8421 \\ \hline
Proposed & \textbf{0.9445} & \textbf{0.9044} & \textbf{0.8379} & \textbf{0.8012} & \textbf{0.8299} & \textbf{0.8636} \\ \hline
\end{tabular}
\end{table}

\vspace{3pt} \noindent \textbf{Performance for DR Staging:} 
Table~\ref{tab:ComparisonAUCs} shows the DR staging AUC for different FL paradigms on different clients.
Our FedUAA achieves the highest AUC scores on all clients, with a 1.48\% improvement in average AUC compared to FedBN~\cite{FedBN}, which achieved the highest average AUC score among the compared methods.
Meanwhile, most FL based approaches achieve higher DR staging performance than SingleSet, suggesting that collaborative training across multiple institutions can improve the performance of DR staging with high data privacy security.
Moreover, as shown in Table~\ref{tab:ComparisonAUCs}, FL paradigms such as FedDyn~\cite{FedDyn} and SCAFFOLD~\cite{SCAFFOLD} exhibit limited performance in our collaborative DR staging task due to the varying staging criteria across different clients, as well as significant differences in label distribution and domain features.
These results indicate that our FedUAA is more effective than other FL methods for collaborative DR staging tasks.
Furthermore, although all FL methods achieve comparable performance on APTOS and DDR clients with distinct features, our FedUAA approach significantly improves performance on clients with small data volumes or large heterogeneity distribution, such as DRR, Messidor, and IDRiD, by 1.27\%, 1.33\%, and 1.29\% over suboptimal results, respectively, which further demonstrates the effectiveness of our core idea of adaptively adjusting aggregation weights based on the reliability of each client.
In addition, we also conduct experiments demonstrate the statistical significance of performance improvement. As shown in Supplementary D, most average p-values are smaller than 0.05. These experimental results further prove the effectiveness of our proposed FedUAA.

\begin{figure}[!t]
\centering
\includegraphics[width=1\textwidth]{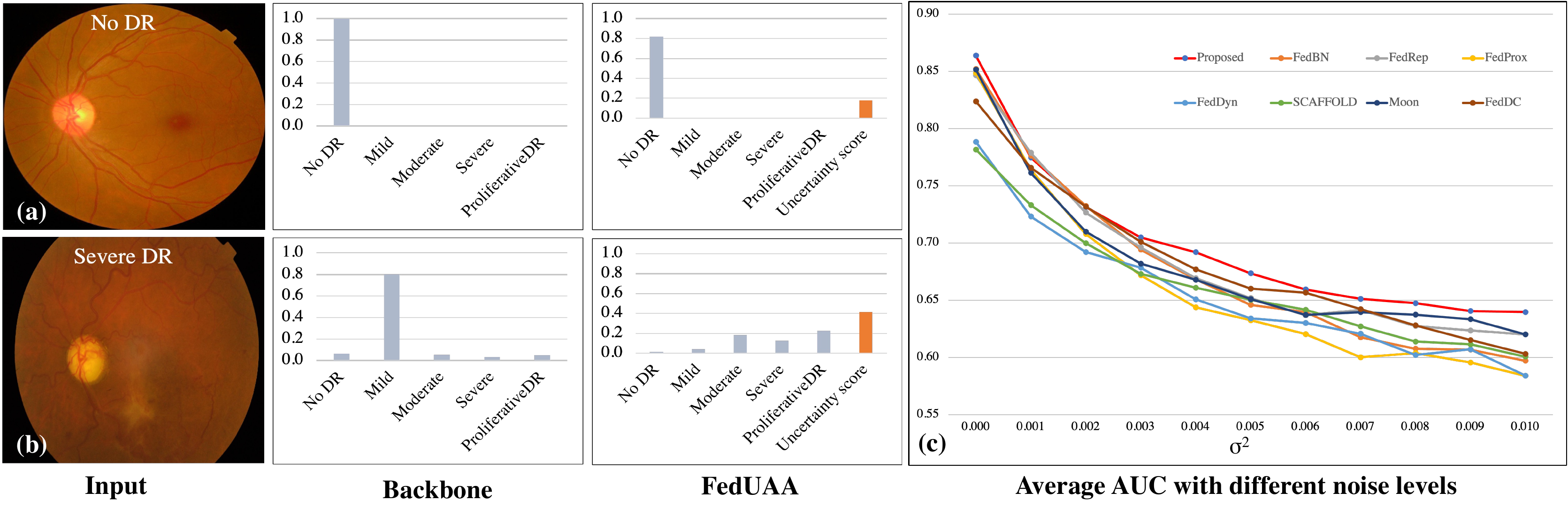}
\vskip -5pt
\caption{(a) Instance of being correctly predicted (b) Sample with incorrect prediction result (c) Average AUC of different methods with increasing noise levels~($\sigma^2$).} 
\label{Reliability}
\end{figure}

\vspace{3pt} \noindent \textbf{Reliability Analysis:}
Providing reliable evaluation for final predictions is crucial for AI models to be deployed in clinical practice.
As illustrated in Fig.~\ref{Reliability}~(b), the model without introducing uncertainty~(Backbone) assigns high probability values for incorrect staging results without any alert messages, 
which is also a significant cause of low user confidence in the deployment of AI models to medical practices.
Interestingly, our FedUAA can evaluate the reliability of the final decision through the uncertainty score.
For example, for the data with obvious features~(Fig.~\ref{Reliability}~(a)), our FedUAA produces a correct prediction result with a low uncertainty score, indicating that the decision is reliable.
Conversely, even if our FedUAA gives an incorrect decision for the data with ambiguous features~(Fig.~\ref{Reliability}~(b)), it can indicate that the diagnosis result may be unreliable by assigning a higher uncertainty score, thus suggesting that the subject should seek a double-check from an ophthalmologist to avoid mis-diagnosis. 
Furthermore, as shown in Fig.~\ref{Reliability}~(c), we degraded the quality of the input image by adding different levels of Gaussian noise $\sigma^{2}$ to further verify the robustness of FedUAA.
Seen from Fig.~\ref{Reliability}~(c), the performance of all methods decreases as the level of added noise increases, however, our FedUAA still maintains a higher performance than other comparison methods, demonstrating the robustness of our FedUAA.

\begin{table}[!t]
\centering
\caption{AUC results for different FL paradigms applied to DR staging.}\label{tab:AblationResults}
\begin{tabular}{l|c|c|c|c|c|c|c|c|c||c}
\hline
Strategy & BC & EU & TWEU & UAW & APTOS & DDR & DRR & Messidor & IDRiD & Average\\ \hline
\multirow{3}{*}{SingleSet} & \cmark & \xmark & \xmark & \xmark & 0.9059 & 0.8776 & 0.8072 & 0.7242 & 0.7168  & 0.8063\\ \cline{2-11} 
 & \cmark & \cmark & \xmark & \xmark & 0.9286 & 0.8589 & 0.8001 & 0.7404 & 0.6928 & 0.8042 \\ \cline{2-11} 
 & \cmark & \xmark & \cmark & \xmark & 0.9414 & 0.8912 & 0.8279 & 0.7309 & 0.7616 & 0.8306 \\ \hline
\multirow{4}{*}{FL} & \cmark & \xmark & \xmark & \xmark & 0.9335 & 0.9003 & 0.8274 & 0.7792 & 0.8193  & 0.8519 \\ \cline{2-11} 
 & \cmark & \cmark & \xmark & \xmark & 0.9330 & 0.8572 & 0.7938 & 0.7860 & 0.7783 & 0.8297\\ \cline{2-11} 
 & \cmark & \xmark & \cmark & \xmark & 0.9445 & 0.8998 & 0.8229 & 0.8002 & 0.8231 & 0.8581\\ \cline{2-11} 
 & \cmark & \xmark & \cmark & \cmark & \textbf{0.9445} & \textbf{0.9044} & \textbf{0.8379} & \textbf{0.8012} & \textbf{0.8299} & \textbf{0.8636}\\ \hline
\end{tabular}
\end{table}

\vspace{3pt} \noindent \textbf{Ablation Study:}
We also conduct ablation experiments to verify the effectiveness of the components in our FedUAA.
In this paper, the pre-trained ResNet50~\cite{ResNet} is adopted as our backbone~(BC) for SingleSet DR staging, while employing FedBN~\cite{FedBN} as the FL BC.
Furthermore, most ensemble-based~\cite{lakshminarayanan2017simple} and MC-dropout-based~\cite{gal2016dropout} uncertainty methods are challenging to extend to our federated DR staging task across multiple institutions with different staging criteria.
Therefore, we compare our proposed method with the commonly used evidential based uncertainty approach~(EU~($L_{Uce}$))~\cite{TMC}. 

For training model with SingleSet, as shown in Table~\ref{tab:AblationResults}, since Dirichlet concentration alters the original feature distribution of the backbone~\cite{TMC}, resulting in a decrease in the model's confidence in category-related evidence, consequently, a decrease in performance when directly introducing EU~(BC+EU~($L_{Uce}$)) for DR staging.
In contrast, our proposed BC+TWEU~($L_{Uce}$+$L_{Tce}$) achieves superior performance compared to BC and BC+EU~($L_{Uce}$), demonstrating that TWEU~($L_{Uce}$+$L_{Tce}$) enables the model to generate a reliable final decision without sacrificing performance.
For training model with FL, as shown in Table~\ref{tab:AblationResults}, BC+FL outperforms SingleSet, indicating that introducing FL can effectively improve the performance for DR staging while maintaining high data privacy security.
Besides, FL+EU~($L_{Uce}$) and FL+TWEU~($L_{Uce}$+$L_{Tce}$) also obtain a similar conclusion as in SingleSet, further proving the effectiveness of TWEU.
Meanwhile, the performance of our FedUAA~(FL+TWEU~($L_{Uce}$+$L_{Tce}$)+UAW) achieves higher performance than FL+TWEU~($L_{Uce}$+$L_{Tce}$) and FL backbone, especially for clients with large data distribution heterogeneity such as DRR, Messidor, and IDRiD.
These results show that our proposed UAW can further improve the performance of FL in collaborative DR staging tasks.
\section{Conclusion}
In this paper, focusing on the challenges in the collaborative DR staging between institutions with different DR staging criteria, we propose a novel FedUAA by combining the FL with evidential uncertainty theory. 
Compared to other FL methods, our FedUAA can produce reliable and robust DR staging results with uncertainty evaluation, and further enhance the collaborative DR staging performance by dynamically aggregating knowledge from different clients based on their reliability.
Comprehensive experimental results show that our FedUAA addresses the challenges in collaborative DR staging across multiple institutions, and achieves a robust and reliable DR staging performance.
\subsubsection{Acknowledgements}
This work was supported by the National Research Foundation, Singapore under its AI Singapore Programme (AISG Award No: AISG2-TC-2021-003), the Agency for Science, Technology and Research (A*STAR) through its AME Programmatic Funding Scheme Under Project A20H4b0141, A*STAR Central Research Fund "A Secure and Privacy Preserving AI Platform for Digital Health”, and A*STAR Career Development Fund (C222812010).

\bibliographystyle{splncs04}
\bibliography{references}

\begin{thebibliography}{10}
\providecommand{\url}[1]{\texttt{#1}}
\providecommand{\urlprefix}{URL }
\providecommand{\doi}[1]{https://doi.org/#1}

\bibitem{FedDyn}
Acar, D.A.E., Zhao, Y., Navarro, R.M., Mattina, M., Whatmough, P.N., Saligrama,
  V.: Federated learning based on dynamic regularization. arXiv preprint
  arXiv:2111.04263  (2021)

\bibitem{arivazhagan2019federated}
Arivazhagan, M.G., Aggarwal, V., Singh, A.K., Choudhary, S.: Federated learning
  with personalization layers. arXiv preprint arXiv:1912.00818  (2019)

\bibitem{asiri2019deep}
Asiri, N., Hussain, M., Al~Adel, F., Alzaidi, N.: Deep learning based
  computer-aided diagnosis systems for diabetic retinopathy: A survey.
  Artificial intelligence in medicine  \textbf{99},  101701 (2019)

\bibitem{FedRep}
Collins, L., Hassani, H., Mokhtari, A., Shakkottai, S.: Exploiting shared
  representations for personalized federated learning. In: International
  Conference on Machine Learning. pp. 2089--2099. PMLR (2021)

\bibitem{connor1969concepts}
Connor, R.J., Mosimann, J.E.: Concepts of independence for proportions with a
  generalization of the dirichlet distribution. Journal of the American
  Statistical Association  \textbf{64}(325),  194--206 (1969)

\bibitem{decenciere2014feedback}
Decenci{\`e}re, E., Zhang, X., Cazuguel, G., et~al.: Feedback on a publicly
  distributed image database: the messidor database. Image Analysis \&
  Stereology  \textbf{33}(3),  231--234 (2014)

\bibitem{fluss2005estimation}
Fluss, R., Faraggi, D., Reiser, B.: {Estimation of the Youden Index and its
  associated cutoff point}. Biometrical Journal: Journal of Mathematical
  Methods in Biosciences  \textbf{47}(4),  458--472 (2005)

\bibitem{gal2016dropout}
Gal, Y., Ghahramani, Z.: Dropout as a bayesian approximation: Representing
  model uncertainty in deep learning. In: international conference on machine
  learning. pp. 1050--1059. PMLR (2016)

\bibitem{FedDC}
Gao, L., Fu, H., Li, L., Chen, Y., Xu, M., Xu, C.Z.: Feddc: Federated learning
  with non-iid data via local drift decoupling and correction. In: CVPR. pp.
  10112--10121 (2022)

\bibitem{Gulshan2016}
Gulshan, V., Peng, L., Coram, M., et~al.: {Development and Validation of a Deep
  Learning Algorithm for Detection of Diabetic Retinopathy in Retinal Fundus
  Photographs}. JAMA  \textbf{316}(22), ~2402 (dec 2016)

\bibitem{gunasekeran2020artificial}
Gunasekeran, D.V., Ting, D.S., Tan, G.S., Wong, T.Y.: Artificial intelligence
  for diabetic retinopathy screening, prediction and management. Current
  opinion in ophthalmology  \textbf{31}(5),  357--365 (2020)

\bibitem{TMC}
Han, Z., Zhang, C., Fu, H., Zhou, J.T.: Trusted multi-view classification.
  arXiv preprint arXiv:2102.02051  (2021)

\bibitem{ResNet}
He, K., Zhang, X., Ren, S., Sun, J.: Deep residual learning for image
  recognition. In: Proceedings of the IEEE conference on computer vision and
  pattern recognition. pp. 770--778 (2016)

\bibitem{huang2022evidence}
Huang, L., Denoeux, T., Vera, P., Ruan, S.: Evidence fusion with contextual
  discounting for multi-modality medical image segmentation. In: MICCAI. pp.
  401--411. Springer (2022)

\bibitem{Kairouz2021}
Kairouz, P., McMahan, H.B., Avent, B., et~al.: {Advances and Open Problems in
  Federated Learning}. Foundations and Trends{\textregistered} in Machine
  Learning  \textbf{14}(1–2),  1--210 (2021). \doi{10.1561/2200000083}

\bibitem{Fed_SCAFFOLD}
Karimireddy, S.P., Kale, S., Mohri, M., Reddi, S., Stich, S., Suresh, A.T.:
  Scaffold: Stochastic controlled averaging for federated learning. In:
  International Conference on Machine Learning. pp. 5132--5143. PMLR (2020)

\bibitem{SCAFFOLD}
Karimireddy, S.P., Kale, S., Mohri, M., Reddi, S., Stich, S., Suresh, A.T.:
  Scaffold: Stochastic controlled averaging for federated learning. In:
  International Conference on Machine Learning. pp. 5132--5143. PMLR (2020)

\bibitem{lakshminarayanan2017simple}
Lakshminarayanan, B., Pritzel, A., Blundell, C.: Simple and scalable predictive
  uncertainty estimation using deep ensembles. Advances in neural information
  processing systems  \textbf{30} (2017)

\bibitem{FedMoon}
Li, Q., He, B., Song, D.: Model-contrastive federated learning. In: Proceedings
  of the IEEE/CVF Conference on Computer Vision and Pattern Recognition. pp.
  10713--10722 (2021)

\bibitem{fundus_survey}
Li, T., Bo, W., Hu, C., Kang, H., Liu, H., Wang, K., Fu, H.: {Applications of
  deep learning in fundus images: A review}. Medical Image Analysis
  \textbf{69},  101971 (apr 2021)

\bibitem{LI2019}
Li, T., Gao, Y., Wang, K., Guo, S., Liu, H., Kang, H.: Diagnostic assessment of
  deep learning algorithms for diabetic retinopathy screening. Information
  Sciences  \textbf{501},  511 -- 522 (2019)

\bibitem{Li2020_FL}
Li, T., Sahu, A.K., Talwalkar, A., Smith, V.: {Federated Learning: Challenges,
  Methods, and Future Directions}. IEEE Signal Processing Magazine
  \textbf{37}(3),  50--60 (may 2020)

\bibitem{FedProx}
Li, T., Sahu, A.K., Zaheer, M., Sanjabi, M., Talwalkar, A., Smith, V.:
  Federated optimization in heterogeneous networks. Proceedings of Machine
  learning and systems  \textbf{2},  429--450 (2020)

\bibitem{FedBN}
Li, X., Jiang, M., Zhang, X., Kamp, M., Dou, Q.: Fedbn: Federated learning on
  non-iid features via local batch normalization. arXiv preprint
  arXiv:2102.07623  (2021)

\bibitem{FedAvg}
McMahan, B., Moore, E., Ramage, D., Hampson, S., y~Arcas, B.A.:
  Communication-efficient learning of deep networks from decentralized data.
  In: Artificial intelligence and statistics. pp. 1273--1282. PMLR (2017)

\bibitem{diagnostics12112835}
Nguyen, T.X., Ran, A.R., Hu, X., Yang, D., Jiang, M., Dou, Q., Cheung, C.Y.:
  Federated learning in ocular imaging: Current progress and future direction.
  Diagnostics  \textbf{12}(11) (2022)

\bibitem{IDRiD}
Porwal, P., Pachade, S., Kamble, R., Kokare, M., Deshmukh, G., Sahasrabuddhe,
  V., Meriaudeau, F.: Indian diabetic retinopathy image dataset (idrid): a
  database for diabetic retinopathy screening research. Data  \textbf{3}(3),
  ~25 (2018)

\bibitem{Ting2017}
Ting, D.S.W., Cheung, C.Y.L., Lim, G., et~al.: {Development and Validation of a
  Deep Learning System for Diabetic Retinopathy and Related Eye Diseases Using
  Retinal Images From Multiethnic Populations With Diabetes}. JAMA
  \textbf{318}(22), ~2211 (dec 2017)

\bibitem{yu2022robust}
Yu, Y., Bates, S., Ma, Y., Jordan, M.: Robust calibration with multi-domain
  temperature scaling. Advances in Neural Information Processing Systems
  \textbf{35},  27510--27523 (2022)

\bibitem{10.1007/978-3-031-16449-1_65}
Zhou, Y., Bai, S., Zhou, T., Zhang, Y., Fu, H.: {Delving into Local Features
  for Open-Set Domain Adaptation in Fundus Image Analysis}. In: MICCAI. pp.
  682--692. Springer Nature Switzerland, Cham (2022)

\bibitem{zou2022tbrats}
Zou, K., Yuan, X., Shen, X., Wang, M., Fu, H.: {TBraTS: Trusted brain tumor
  segmentation}. In: MICCAI. pp. 503--513. Springer (2022)

\end{thebibliography}
%





\newpage
\begin{appendix}
\section*{Supplementary Materials}
\section*{A. Algorithm of our proposed FedUAA}
\begin{algorithm}[h]
\caption{Collaborative DR staging using FedUAA.}\label{alg:alg1}
\begin{algorithmic}[1]
\REQUIRE{Datasets from $N$ clients: $D_{1},...,D_{N}$; total optimization round $T$; initialize $\Phi =\left[ \varphi_{1} ,...,\varphi_{N} \right]$, $\Psi =\left[ \psi_{1} ,...,\psi_{N} \right]$; Prediction results $P$; Uncertainty distribution $U$; the model of \textit{i}-th client of $f_{i}$,}
\STATE $ \textbf{For \(t=1,..,T\) do} $
\STATE $  \text{\qquad Server sends global encoder $\varphi^{t}_{g} $ to each client.}$
\STATE $  \text{\qquad \textit{\textbf{\textcolor{red}{Local:}}}}$
\STATE $\qquad \textbf{  For each \(client i=1,...,N\) do}$
\STATE $ \text{\qquad \qquad $\varphi^{t}_{i} \leftarrow \varphi^{t}_{g}$.}$
\STATE $ \text{\qquad \qquad
\textbf{Get prediction and uncertainty scores: }$P^{t}_{i},U^{t}_{i}=f^{t}_{i}\left( \varphi^{t}_{i} ,\psi^{t}_{i} |X_{i}\right)$}.$
\STATE $ \text{\qquad \qquad \textbf{Local updates: }$\varphi^{t+1}_{i} ,\psi^{t+1}_{i} \leftarrow SGD\left( f^{t}_{i}\left( \varphi^{t}_{i} ,\psi^{t}_{i} |X_{i}\right)  \right)$}.$
\STATE $ \text{\qquad \qquad \textbf{1.}Calculate ground truth for mis-prediction $U^{GT,t}_{i}$ by Eq.~2~(Sec 2.2).} $ 
\STATE $ \text{\qquad \qquad \textbf{2. }Find the optimal uncertainty score $\theta^{t}_{i}$ that can explicitly reflect the
}$
\STATE $ \text{\qquad \qquad reliability of $clienti$ by Eq.~3~(Sec 2.2).
}$
\STATE $ \textbf{\qquad End For} $
\STATE $ \text{\qquad Send $\Theta^{t} =\left[ \theta^{t}_{1} ,...,\theta^{t}_{N} \right] $ and $\Phi^{t+1} =\left[ \varphi^{t+1}_{1} ,...,\varphi^{t+1}_{N} \right]$ to server.}$ 
\STATE $  \text{\qquad \textit{\textbf{\textcolor{red}{Server:}}}}$
\STATE $ \text{\qquad  Normalize $\Theta^{t}$ to obtain model aggregation weights $w^{t+1}_{i}={e^{\theta^{t}_{i} }}  /{\sum\nolimits^{N}_{i=1} e^{\theta^{t}_{i} }}$}.$ 
\STATE $ \text{\qquad \textbf{Model aggregation:} $\varphi^{t+1}_{g} =\sum^{N}_{i=1} w^{t+1}_{i}\ast \varphi^{t+1}_{i}$}$ 
\STATE $ \textbf{End For} $
\end{algorithmic}
\label{alg1}
\end{algorithm}

\section*{B. Youden index}
The Youden Index is a statistic used to evaluate the performance of a binary classification model. It takes into account both the sensitivity and specificity of the model and is defined as:
\begin{equation} 
J=Sensitivity+Specificity-1.
\end{equation}
The Youden Index considers both the ability of the test to correctly identify positive cases (sensitivity) and negative cases (specificity) and is therefore less likely to be biased towards either true positives or true negatives. And, The Youden Index ranges from 0 to 1, with a value of 0 indicating that the model has no discriminative ability and a value of 1 indicating perfect discriminative ability.
In this paper, we evaluate the reliability of the model by calculating the consistency between the uncertainty score distribution~($U$) and the ground truth of mis-prediction~($U^{GT}$) obtained by Eq.~2~(Sec.2.2).
This process can be seen as a binary classification problem, and the closer $U$ is to $U^{GT}$, the higher the model reliability.
To find the optimal uncertainty score that can well reflect the model's reliability, we calculate the ROC curve between $U$ and $U^{GT}$, and obtain all possible sensitivity~($Sens$) and specificity~($Spes$) values corresponding to each uncertainty score~($u$) used as a threshold. Therefore, based on Youden Index, the final optimal uncertainty score $\theta$ can be obtained by Eq.~3~(Sec.2.2).

\section*{C. Details of datasets}
We construct a database for federated DR staging based on 5 public datasets, including APTOS (3,662 samples), Messidor (1,200 samples), DDR (13,673 samples), KaggleDR (35,126 samples) (DRR), and IDRiD (516 samples), where each dataset is regarded as a client, More details of datasets are shown in Table~\ref{tab:DataDetails}
\begin{table}[h]
\centering
\caption{Details for different datasets.}
\label{tab:DataDetails}
\begin{tabular}{l|l|l|l|l|l}
\hline
Datasets & DR staging criteria & Devices & Train & Validation & Test \\ \hline
APTOS & Normal+4 stages & Multiple devices & 2,930 & 366 & 366 \\ \hline
Messidor & Normal+3 stages & Topcon TRC NW6 & 718 & 240 & 242 \\ \hline
DDR & Normal+4 stages+poor quality & 42 types of devices & 6,835 & 2,733 & 4,105 \\ \hline
DRR & Normal+4 stages & Various devices & 21,074 & 7,026 & 7,026 \\ \hline
IDRiD & Normal+4 stages & Kowa VX-10 & 329 & 84 & 103 \\ \hline
\end{tabular}
\end{table}
\section*{D. Statistical significance of performance improvement}
To demonstrate the statistical significance of performance improvement, we further perform 3-trial repeating experiment with different random seeds and calculate average p-value between the proposed method and other comparison baselines. As shown in Table~\ref{tab:P_Values}, most average p-values are smaller than 0.05. These experimental results further prove the effectiveness of our proposed FedUAA.
\begin{table}[h]
\centering
\caption{P-value for comparing the performance improvement of different methods.}\label{tab:P_Values}
\begin{tabular}{l|c|c|c|c|c|c}
\hline
Methods & APTOS & DDR & DRR & Messidor & IDRiD & Average \\ \hline
Proposed-SingleSet & 0.046 & 0.063 & 0.141 & 0.111 & 0.076 & 0.001 \\ \hline
Proposed-FedRep & 0.177 & 0.093 & 0.125 & 0.042 & 0.173 & 0.003 \\ \hline
Proposed-FedBN & 0.133 & 0.137 & 0.252 & 0.171 & 0.643 & 0.000 \\ \hline
Proposed-FedProx & 0.059 & 0.123 & 0.050 & 0.041 & 0.049 & 0.005 \\ \hline
Proposed-FedDyn & 0.018 & 0.020 & 0.003 & 0.009 & 0.007 & 0.007 \\ \hline
Proposed-FedDC & 0.192 & 0.030 & 0.428 & 0.010 & 0.054 & 0.008 \\ \hline
Proposed-Moon & 0.479 & 0.289 & 0.023 & 0.014 & 0.312 & 0.051 \\ \hline
Proposed-SCAFFOLD & 0.015 & 0.019 & 0.088 & 0.075 & 0.002 & 0.008 \\ \hline
Proposed-MDT & 0.016 & 0.250 & 0.010 & 0.831 & 0.516 & 0.025 \\ \hline
\end{tabular}
\end{table}
\end{appendix}

\end{document}